\documentclass[letter]{aa}  

\usepackage{newtxtext,newtxmath}

\usepackage[T1]{fontenc}
\usepackage{ae,aecompl}
\usepackage{comment}

\usepackage{aas_macros}
\usepackage[hidelinks]{hyperref}

\usepackage{graphicx}	
\usepackage[export]{adjustbox}

\usepackage{xcolor}

\def\gsim{\;\raise0.3ex\hbox{$>$\kern-0.75em\raise-1.1ex\hbox{$\sim$}}\;}
\def\Msun{M_\odot}

\def\ergs{\rm ~erg~s^{-1}}
\def\lsim{\;\raise0.3ex\hbox{$<$\kern-0.75em\raise-1.1ex\hbox{$\sim$}}\;}
\def\gsim{\;\raise0.3ex\hbox{$>$\kern-0.75em\raise-1.1ex\hbox{$\sim$}}\;}
\begin{document}

\title{North Polar Spur : gaseous plume(s) from star-forming regions at $\sim$3-5~kpc from Galactic Center?}
\author{
Eugene~Churazov \inst{1,2} 
\and
Ildar~I.~Khabibullin \inst{3,1,2} 
\and
Andrei~M.~Bykov \inst{4} 
\and
Nikolai~N.~Chugai \inst{5} 
\and
Rashid~A.~Sunyaev \inst{2,1} 
\and
Victor~P.~Utrobin \inst{5,6} 
\and
Igor~I.~Zinchenko \inst{7}
} 

\institute{
Max Planck Institute for Astrophysics, Karl-Schwarzschild-Str. 1, D-85741 Garching, Germany 
\and 
Space Research Institute (IKI), Profsoyuznaya 84/32, Moscow 117997, Russia
\and
Universitäts-Sternwarte, Fakultät für Physik, Ludwig-Maximilians-Universität München, Scheinerstr.1, 81679 München, Germany
\and
Ioffe Institute, Politekhnicheskaya st. 26, Saint Petersburg 194021, Russia
\and
Institute of Astronomy, Russian Academy of Sciences, 48 Pyatnitskaya str., Moscow 119017, Russia
\and
NRC `Kurchatov Institute', acad. Kurchatov Square 1, Moscow 123182, Russia
\and
Institute of Applied Physics of the Russian Academy of Sciences, 46 Ul'yanov~str., Nizhny Novgorod 603950, Russia
}



\abstract{
We argue that the North Polar Spur (NPS) and many less prominent structures are formed by gaseous metal-rich plumes associated with star-forming regions (SFRs). The SFRs located at the tangent to the 3-5~kpc rings might be particularly relevant to NPS.  A multi-temperature mixture of gaseous components and cosmic rays rises above the Galactic disk under the action of their initial momentum and buoyancy. 
Eventually, the plume velocity becomes equal to that of the ambient gas, which rotates with different angular speed than the stars in the disk.  As a result, the plumes acquire characteristic bent shapes. 
An ad hoc model of plumes' trajectories shows an interesting 
resemblance to the morphology of structures seen in the radio continuum and X-rays. 
}

\titlerunning{North Polar Spur}

    
\maketitle

\section{Introduction}
\label{sec:intro}


North Polar Spur \citep[NPS,][]{1958PMag....3..370T} is an elongated bright structure first discovered in all-sky radio images along with other similar structures dubbed Loops \citep[e.g. ][ for a recent review]{2018Galax...6...56D}. Unlike other Loops, NPS (the northern part of Loop~I) is visible in soft X-rays across its full extent \citep{1972ApJ...172L..67B,1995A&A...294L..25E,2018Galax...6...27K,2020Natur.588..227P,2020ApJ...904...54L,2021ApJ...908...14K}.    

Given the NPS's large angular size, $\gtrsim90\deg$, the first proposed explanation \citep[e.g.][]{1960Obs....80..191H,1971A&A....14..359B} was a shell of a very nearby, $d\lesssim100$ pc, supernova remnant. Alternatively, a scenario in which NPS represents a distant, $d\gtrsim10$ kpc, Galaxy-scale structure has been put forward, with the possible energetic outburst in the Galactic center being considered as the primary source behind it \citep{1977A&A....60..327S,1994ApJ...431L..91S}. The latter might be caused by an intense star-formation episode in the vicinity of the Galactic Center or by the activity of the supermassive black hole Sgr~A* \citep[e.g.][]{1977A&A....60..327S,1994ApJ...431L..91S,2015MNRAS.453.3827S,2021MNRAS.506.2170S,2022NatAs...6..584Y}. Both scenarios suggest that NPS is powered by a shock wave propagating in the hot medium (either interstellar or circumgalactic). This shock is responsible for the acceleration of radio-emitting relativistic electrons, as well as the heating and compression of the X-ray-emitting hot gas. 

Although these two scenarios involve very different physical sizes, time scales and required energetics, it turned out to be very difficult to unequivocally distinguish between them based on the available data, with contradicting conclusions having been reached from, e.g., the Faraday Rotation and X-ray absorption measurements \citep[see a comprehensive discussion in ][]{2023CRPhy..23S...1L}. Recent discoveries of the giant structures in gamma-rays \citep[Fermi bubbles, ][]{2010ApJ...724.1044S} and soft X-rays in the southern Galactic sky \citep[eROSITA,][]{2020Natur.588..227P}, the "bubbles" appear to be more symmetric with respect to the Galactic center, strengthening the case for the central energy release as a viable solution \citep[e.g.][]{2024A&ARv..32....1S}.  

In this Letter, we consider another scenario for NPS formation motivated by the morphological and spectral properties of the soft X-ray emission measured in the course of the eROSITA all-sky survey \citep{2021A&A...647A...1P,2021A&A...656A.132S}. We propose that NPS is produced by a break-out of the massive star formation regions associated with the end of the Galactic bar, which locates its base at $d\sim5$ kpc from us \citep[e.g., ][]{2016ARA&A..54..529B}. Qualitatively similar arguments have recently been discussed in the work of \cite{2024ApJ...973...78S}.  Flows of enriched hot gas move upward from the Galactic disk and get entrained in the relative rotation of the circumgalactic medium above the disk, resulting in the spiral-like structure, appearing like a loop in the sky. In this model, it is the advection of the enriched gas and relativistic particles, rather than a shock, that is responsible for the appearance of NPS, predicting the high metal abundance of X-ray emitting gas and lack of evolutionary signatures in the direction perpendicular to the NPS edge.


\begin{figure*}
\centering
\includegraphics[angle=0,trim=1.6cm 7.3cm 1.6cm 7.3cm,width=1.8\columnwidth]{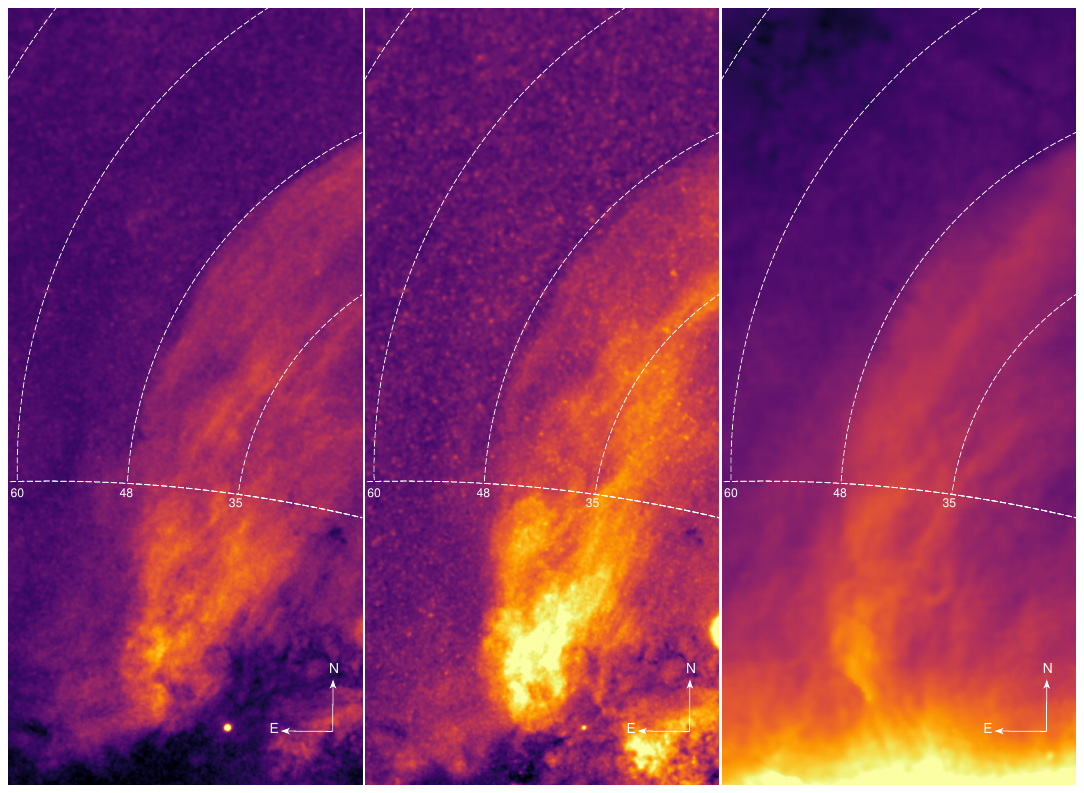}
\caption{X-ray and radio images of the NPS region. The left and middle panels:  the 0.52-0.61 keV and 0.7-0.96 keV eROSITA X-ray images, respectively. For comparison, the right panel shows the 408 MHz radio image from the all-sky continuum surveys \citep{1981A&A...100..209H,2015MNRAS.451.4311R}.  The images in galactic coordinates are shown in stereographic projection for a better view of the regions near the Galactic plane and the Galactic North Pole. The X-ray images are particle background subtracted, and exposure corrected. The brightest compact sources and galaxy clusters have been masked and the resulting image convolved with $\sigma=20'$ Gaussian. Still prominently visible in the image are the RS Oph (bright dot near the bottom of the left image) and the stray light halo around Sco X-1 (at the right edge of the middle image).  
The white dashed lines show the wedge used for the extraction of radial profiles. The numbers indicate the distance (in degrees) from the wedge center, which is at $(l,b)=(338^\circ,	32^\circ)$ in Galactic coordinates.} 
\label{fig:image_o7fe}
\end{figure*}

\section{X-ray picture}
\label{s:x-ray}

For X-ray analysis, we use the data accumulated by eROSITA \citep{2021A&A...647A...1P} telescope onboard the SRG~ X-ray observatory \citep{2021A&A...656A.132S} in the course of its all-sky survey in the Eastern Galactic hemisphere (i.e. the Galactic longitude range $0^\circ<l<180^\circ$). Details on the data preparation and analysis are given in Appendix~\ref{app:xraydata}. Images extracted from the eROSITA all-sky survey in the 0.52-0.61 keV and 0.7-0.96 keV bands are shown in Fig.~\ref{fig:image_o7fe} (left and middle panels, respectively). The former band includes the triplet of He-like oxygen O~VII near 574~eV, while the bright lines of helium-like neon Ne~IX (at 905 and 921 eV) and neon-like iron Fe~XVII (between 725 and 826 eV) fall into the latter. For plasma in collisional ionization equilibrium (CIE), the peak line emissivities in these bands are at temperatures $\sim 0.2\,{\rm keV}$ and  $\sim 0.6\,{\rm keV}$, respectively.

Rather complex X-ray morphology of NPS is obvious from these images and the comparison between them. One can readily see that the correlation length of X-ray structures is larger in the direction along NPS for both bands but a simple combination of concentric spherical shells is not a good description of NPS.
For example, the surface brightness in the 0.52-0.61 keV band dominated by the O~VII line, has a sharp outer edge but a rather flat surface brightness across the inner region, unlike an edge-brightened shell. Similarly, the 0.7-0.96 keV surface brightness is composed of a fainter diffuse emission surrounding a bright central "filament". 

\begin{figure*}
\centering
\includegraphics[angle=0,trim=1cm 5.5cm 1cm 2.5cm,width=0.95\columnwidth]{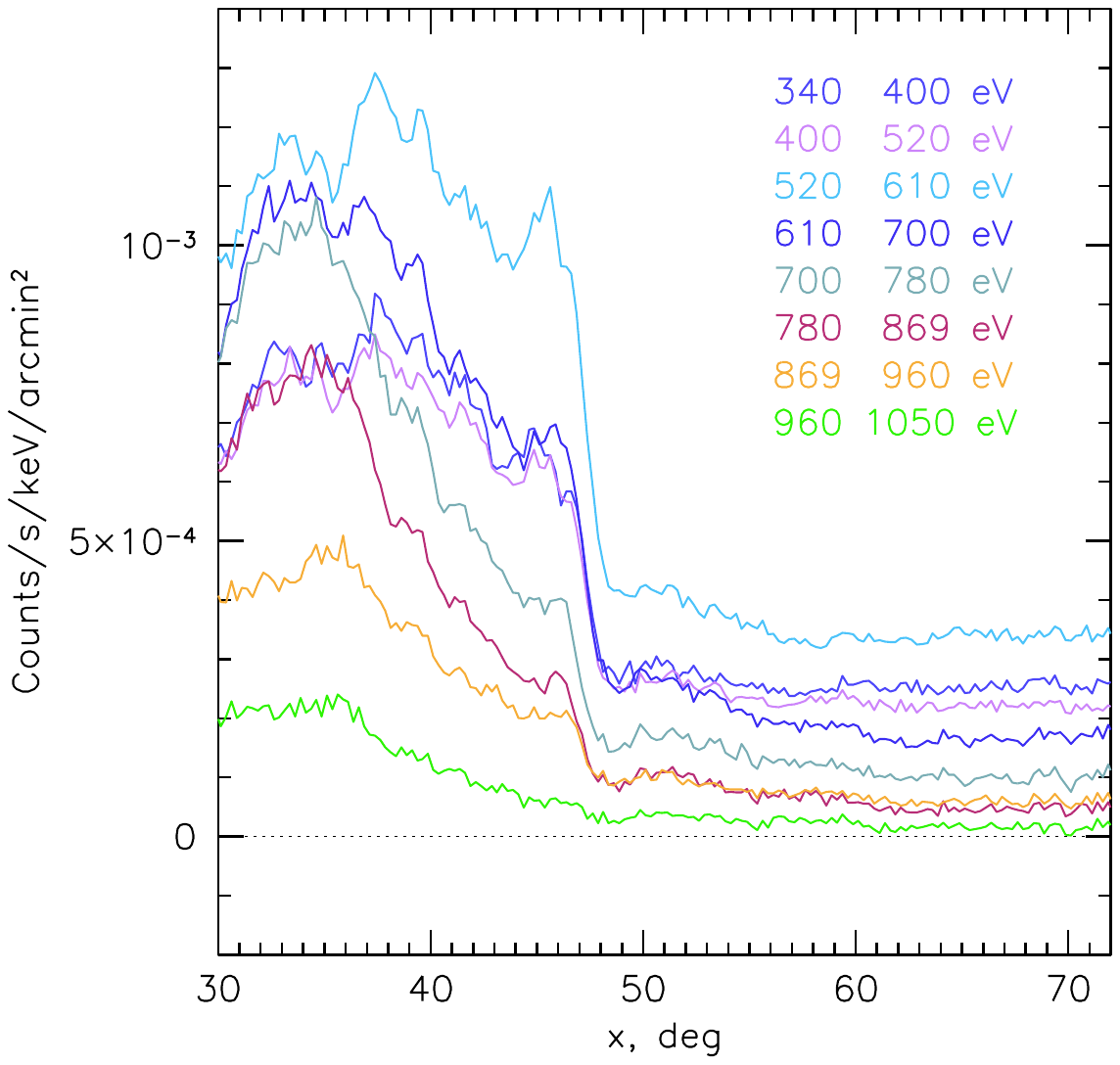}
\includegraphics[angle=0,trim=1cm 5.5cm 1cm 2.5cm,width=0.95\columnwidth]{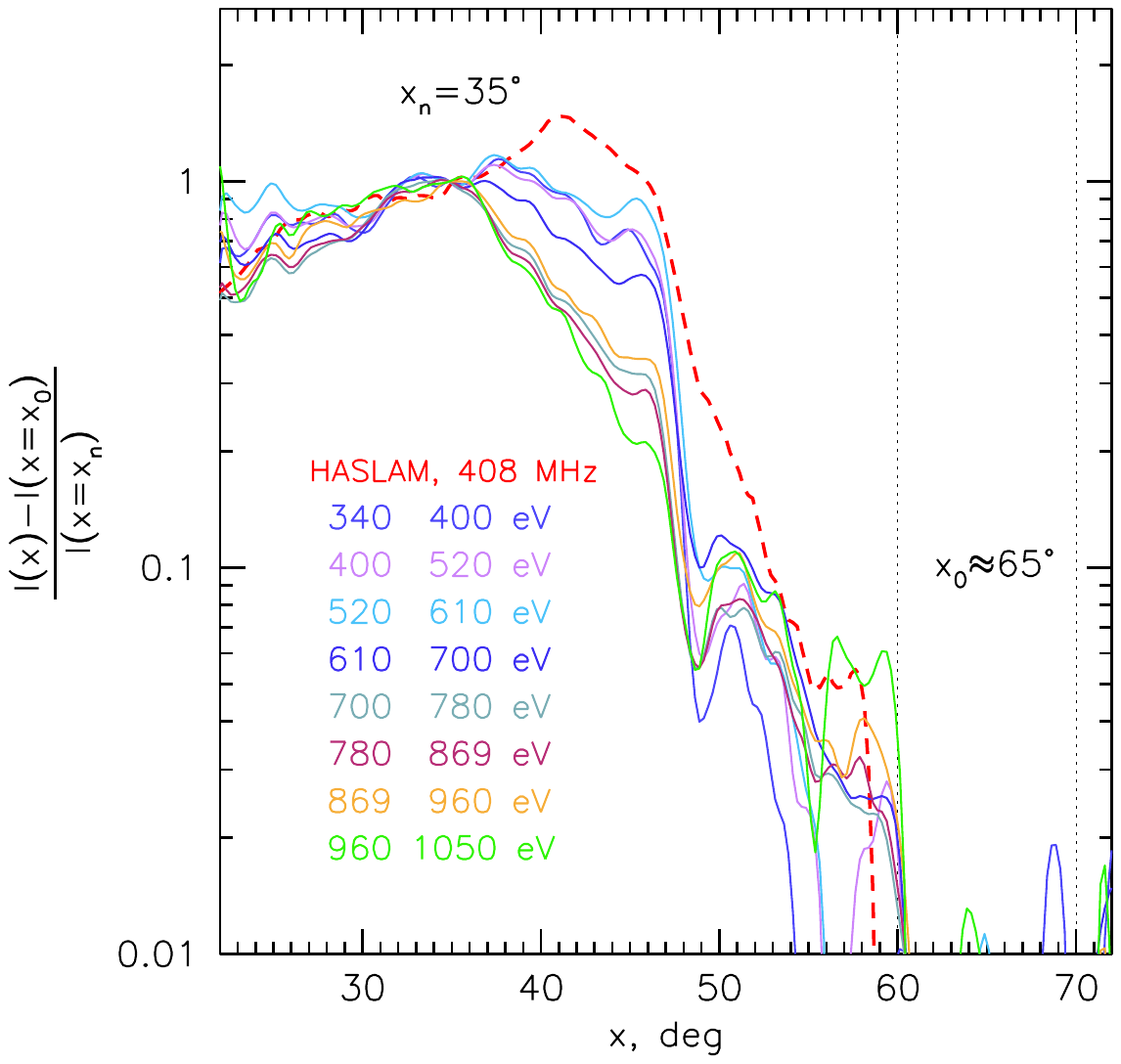}
\caption{{\bf Left:} Radial profiles of the NPS X-ray emission in several energy bands. The wedge used for the flux extraction is shown in Fig.~\ref{fig:image_o7fe}. The bright edge of the NPS is at $x\approx 48\,{\rm degrees}$. The contributions of the instrumental background and distant (unresolved) sources (CXB) have been subtracted, leaving only the flux from the Galaxy. 
{\bf Right:} The same profiles as in the left plot, but after subtracting the mean level at $x=60-70$~degrees, renormalizing by the flux at $x=35$~degrees, and lightly smoothing with a Gaussian filter ($\sigma=0.4^\circ$). In addition, the radio surface brightness profile (Haslam, 408~MHz) is shown with the red dashed curve after applying the same procedure (subtraction, renormalization, and smoothing) as used for the X-ray profiles. 
}
\label{fig:rprof}
\end{figure*}

More quantitatively this is illustrated via profiles of the X-ray emission in several narrow bands. The white dashed lines in Fig.~\ref{fig:image_o7fe} outline a wedge that is aligned with the NPS "outer" boundary. Radial profiles extracted from this wedge are shown in Fig.~\ref{fig:rprof}. The steep rise, corresponding to the "edge" of NPS, is visible in all bands. However, further inside NPS, the profiles do not follow a typical edge-brightened behavior characteristic for a spherical shock associated with an instantaneous point-like energy release in a uniform medium (see, e.g., profiles at energies below 0.61 keV). The observed profiles could be affected by the upstream gas density gradients, Non-Equlibrium Ionization (NEI) effects downstream of the shock, and, especially, by the energy release mode. For example, in a “steady wind” scenario \citep[see, e.g.][]{2015MNRAS.453.3827S}, the surface brightness profile is rather flat.

In the model without the shock that we pursue here, this behavior suggests that the volume occupied by the plasma emitting at these energies resembles a flattened sheet rather than a spherical shell.
 In harder bands, the profiles feature a steep jump, followed by a gradual increase of intensity before reaching a peak at $x\approx 35\,{\rm degrees}$ in this plot. The center of the wedge is at $(l,b)=(338^\circ,32^\circ)$ in Galactic coordinates.

An alternative to the local SNR or GC-related scenarios, which associate the steep rise with a shock, is the assumption that NPS is a multi-temperature gas, where a hotter ($>$0.3 keV) component, responsible for the emission in iron lines above 0.7 keV, coexists with a warm (e.g., 0.15-0.2 keV) plasma, producing the bulk of the emission below 0.7 keV and most prominently in oxygen lines.

The sharpness of the NPS edge is considered as a strong argument in favor of the shock scenario \citep[e.g.][]{2020Natur.588..227P}. However, there are examples of astrophysical objects when very sharp edges are seen in the hot X-ray-emitting plasma. The most relevant are “cold fronts” in galaxy clusters \citep[see][for reviews]{2007PhR...443....1M,2016JPlPh..82c5301Z}. These cold fronts are contact discontinuities rather than shocks, which might retain their sharpness due to magnetic fields and/or accelerated gas motions along the front \citep[see, e.g.][]{2004MNRAS.350L..52C}. Similar effects might be relevant to the rising plumes discussed here. 

Accurate temperature and abundance measurements for gas with temperature below 1 keV (when pure bremsstrahlung continuum is very weak and subdominant to other continuum components) are highly model-dependent. As is clear from Fig.2, the ratio of fluxes in different bands varies strongly across NPS. This means that the best-fitting temperature strongly depends on the region used for spectrum extraction. It further implies that a mixture of components with different temperatures (and different ionization states, especially in the shock-based scenario) will inevitably be present. The abundance determination for the gas in the relevant temperature range (0.1-0.6 keV) is highly degenerate with the emission measure even when the gas is in the Collisional Ionization Equilibrium (CIE). In practice, it is difficult to measure the abundance if it is larger than ~0.1 Solar. The effects of non-equilibrium ionization that are expected in the shock scenario, further affect the “apparent” temperature and abundance. These issues are likely the reason for different temperature/abundance measurements available in the literature \citep[e.g.][]{2013ApJ...779...57K,2022MNRAS.512.2034Y} for various patches of the NPS.  The exceptionally uniform and sensitive eROSITA all-sky data give a clear view of the amplitude of spectral variations across such large regions for the first time. We defer the extended spectral analysis of the NPS to future work.

In what follows, we consider a scenario in which such multi-temperature gas is venting from the active star-forming regions in the central parts of the Galactic disk. 

\begin{figure*}
\centering
\includegraphics[angle=0,trim=1.5cm 9.5cm 1.5cm 9.5cm,clip,width=1.99\columnwidth]{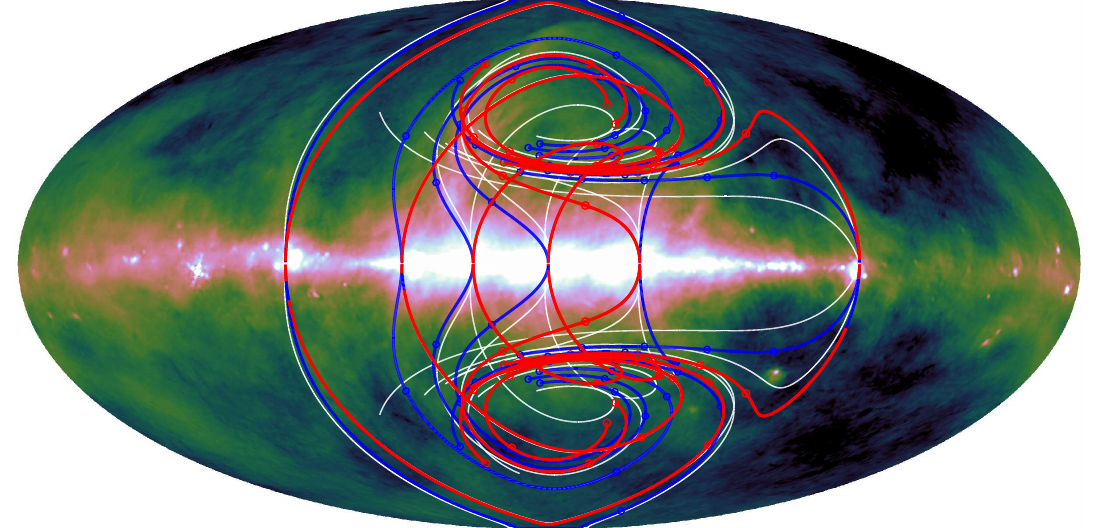}
\includegraphics[angle=0,trim=1cm 9.cm 1cm 9.cm,clip,width=1.99\columnwidth]{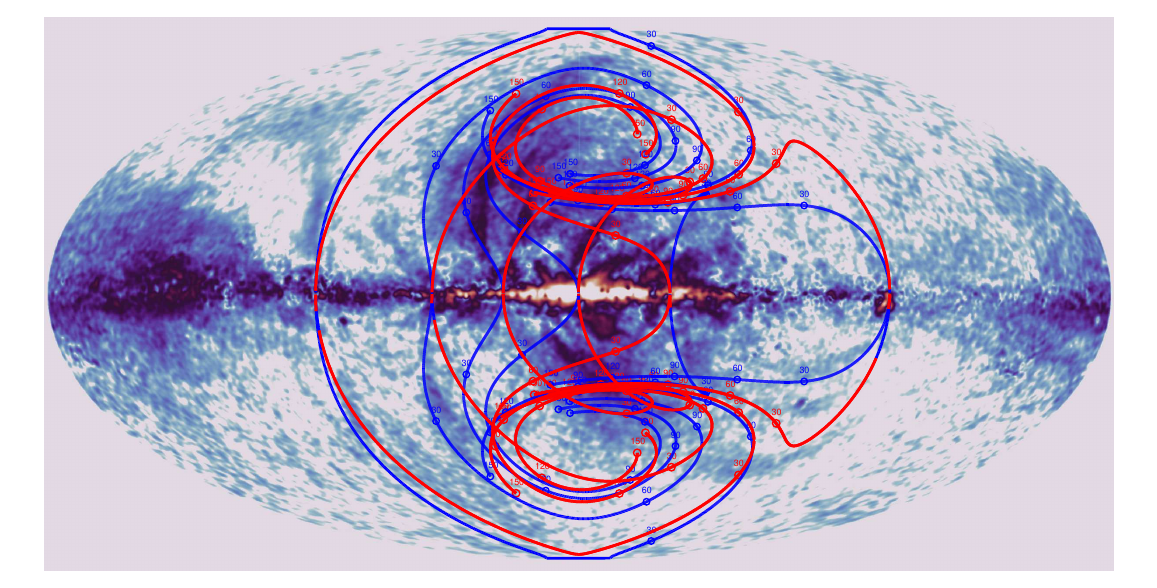}
\caption{{\bf Top.} Sample trajectories of the gaseous plumes superposed on the radio continuum map at 408~MHz map \citep{1981A&A...100..209H,2015MNRAS.451.4311R}. These trajectories correspond to an arbitrarily chosen set of plume "sources" (see Table~\ref{tab:src}). The red and blue colors correspond to cases when sources have the same azimuthal velocity in the disk as stars. They correspond to the plume sources currently observed at a given Galactic longitude but different distances from the Sun: the red and blue colors mark sources that are closer and further away, respectively. The white color corresponds to sources that move with the pattern speed. Three properties of the trajectories are worth mentioning: (i) the form "spirals" winded in the central region of the Galaxy well above the plane, (ii) East-West asymmetry set by the Galaxy rotation direction, and (iii) trajectories tend to overlap in certain regions of the sky even if they come from well-separated sources. {\bf Bottom.} Polarized synchrotron emission map \citep[e.g.][]{2020A&A...641A...4P} with a subset of trajectories superposed. The circles correspond to time tags (every 30~Myr) in this model. In the bottom panel, these tags are labeled.}
\label{fig:traj_sim_radio}
\end{figure*}

\begin{figure}
\centering
\includegraphics[angle=0,trim=1cm 5.5cm 1cm 2.5cm,clip,width=0.95\columnwidth]{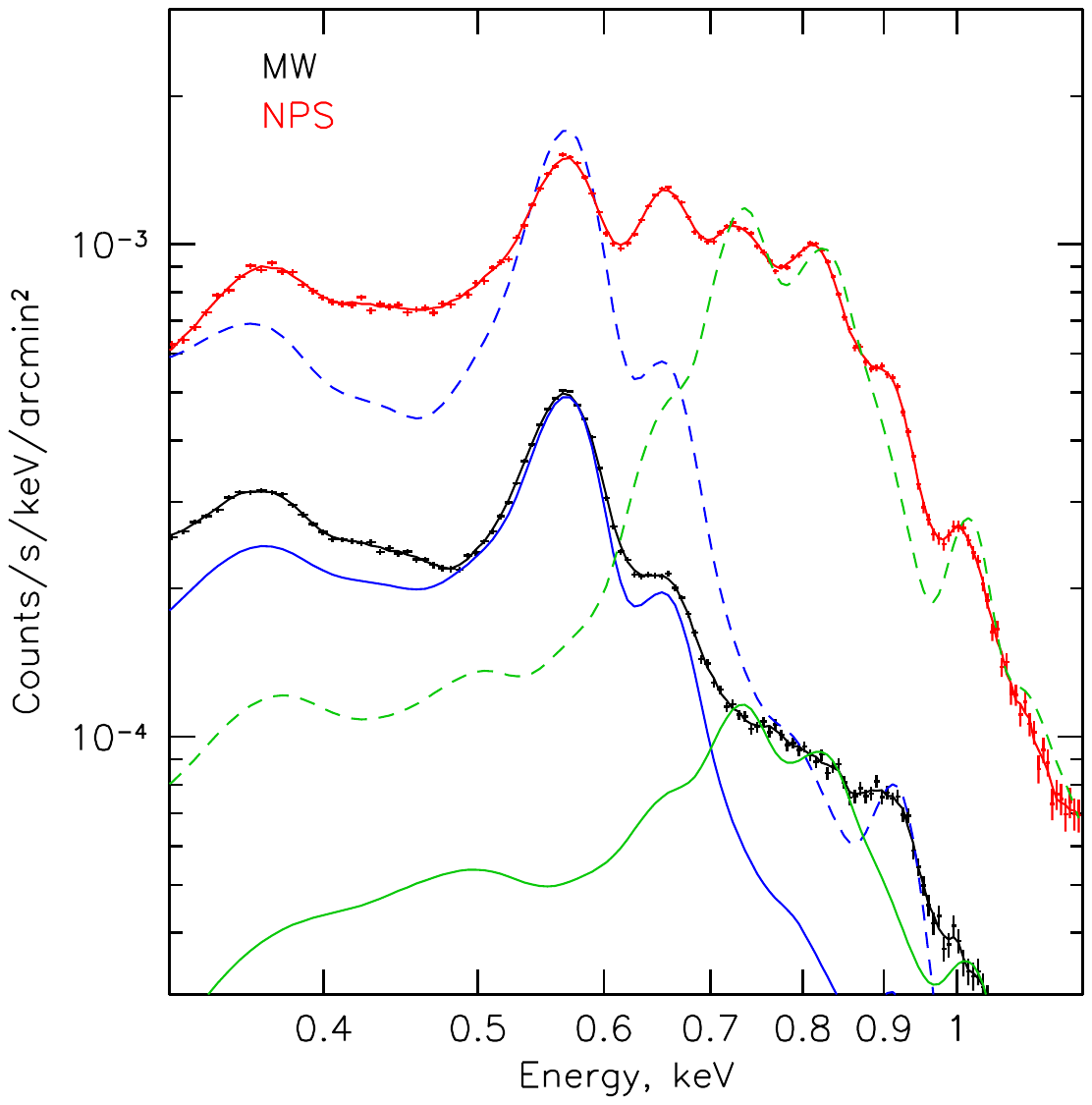}
\caption{
The spectrum of a large region inside NPS (red points) in comparison with the typical "Milky Way" spectrum well outside NPS (black points). Contributions of the detector background and CXB have been subtracted. The blue and green lines illustrate a few characteristic models. The solid blue line shows the APEC spectrum with the temperature $T_{\rm w}=0.16$~keV and abundance $Z/Z_\odot=0.05$, using for the abundance ratios from \cite{2009ARA&A..47..481A}.
The blue dashed line shows the same model with 3 times larger abundance, i.e. $Z/Z_\odot=0.15$. The green solid line shows the APEC model with the temperature $T_{\rm h}=0.5$~keV, $Z/Z_\odot=0.05$, and the emission measure multiplied by factor $(T_{\rm w}/T_{\rm h})^2$. The dashed green line shows the same model with $Z/Z_\odot=0.7$. A comparison of the histograms and the dashed lines shows that over-abundances in the range of 3-10 are needed to reproduce enhanced brightness of the NPS compared to the Galaxy (assuming comparable pressures and linear sizes of emitting regions).}
\label{fig:spec}
\end{figure}

\section{Morphological model}
\label{s:model}

Here we consider a simple morphological model that is based on a broad-brush picture of the evolution of a massive star-forming region in the Galactic disk \citep[e.g.][]{1989ApJ...345..372N,2020ApJ...900...61K}. The first shock waves of supernova explosions originating from the most massive newborn stars in such a region produce hot and dense gas strongly enriched with metals of the ejecta. Initially confined by the surrounding dense and cold gas, the hot gas manages to break through and form an (adiabatically cooled) plume that moves away from the disk and interacts with the interstellar (ISM) and circumgalactic (CGM) media forming some sort of chimneys and plumes (see Appendix~\ref{app:chimney}). The trajectories of rising plumes might be complicated, especially when the gas is multi-phase and magnetized. They are also affected by the pressure, magnetic fields, and velocities in the ambient ISM and/or CGM \citep[e.g. ][]{2023ARA&A..61..131F}.

For illustration purposes, we consider an ad hoc model that encapsulates all the complicated physics in three parameters. Namely, the initial (vertical) velocity of the plume $\varv_{z,0}$ and two effective scale heights $z_\phi$ and $z_z$. Here $z_\phi$ describes how the plume that initially moves together with stars and gas close to the disk plane, decelerates and joins the pressure-supported hot halo gas, which is either non-rotating or rotating slower than the stellar disk. In turn, the $z_z$ parameter controls the decline of the vertical (perpendicular to the disk) velocity component. As a result, the two velocity components have the following dependence on $z$:     

\begin{equation}
\varv_z(z)=\varv_{z,0} \, e^{-z/z_z} ~~~{\rm and}~~
\varv_{\phi}(z)=\varv_{\rm rot} \, \left [1- e^{-z/z_\phi} \right ].
\label{eq:vz}
\end{equation}

A flat rotation curve with $\varv_{\rm rot} = 220 \,{\rm km\,s^{-1}}$ is adopted for the sake of simplicity, the effective scale heights are set to $z_\phi=2\,{\rm kpc}$ and $z_z=5\,{\rm kpc}$, and $\varv_{z,0}= 200 \,{\rm km\,s^{-1}}$. The choice of these parameters is rather arbitrary, but it suffices for illustration purposes and the final answer is not dramatically sensitive to it.

For the source of the plume, e.g. a star-forming region, we assume a circular trajectory in the disk plane with the velocity $\varv_{\phi,\rm src}$. In our illustrative model, the following two versions are considered:
\begin{equation}
({\rm a}) \,\, \varv_{\phi,\rm src}=\varv_{\rm rot}  ~~{\rm and \,\, (b)\,\,} \varv_{\phi,\rm src}=\Omega R  ,
\label{eq:vphi}
\end{equation}
where $R$ is the distance from the Galactic Center. Version (a) corresponds to the source moving with the same velocity as stars, while version (b) mimics a pattern motion. The former case is relevant when the active phase of a given star-forming region associated with the dense gas is very long and the active region always moves together with the gas (with the velocity $\varv_{\rm rot}$). The latter case corresponds to a situation when the spatial distribution of the plume sources is linked to the Galactic bar and/or spiral arms. For illustration of this case, we adopted the value $\Omega=33\,{\rm km\,s^{-1}\,kpc^{-1}}$ derived for the bar pattern speed \citep[e.g.][]{2022MNRAS.512.2171C}. 
As a further simplification, we assume that none of these parameters depend on the position of the plume within the galaxy. 

The above relations (\ref{eq:vz}-\ref{eq:vphi}) can be trivially integrated to derive a position of the gas lump released some time $t$ ago relative to the current position of the "source". By design, plume trajectories appear in 3D as spirals rising from the disk from the current location of the plume source (see Appendix~\ref{app:plumes} and Fig.~\ref{fig:3d}). 
We note here that for a strongly underdense lump, the buoyancy force is solely set by pressure gradients in the ambient gas. These gradients themselves depend on the rotation velocity of the halo gas and can have both radial and vertical components. In the above model, we explicitly neglect the former component, although in real systems it can force the plume to move radially. In particular, for slow (fast) rotation of the halo gas, underdense lumps can move to larger (smaller) radii (see Fig.~\ref{fig:3dbub}).

As a final step, we project the trajectory onto the sky as seen from the position of the Sun. This modifies the appearance of the plume due to different distances from the Sun to various segments of the plume 
and the Sun's motion relative to the source. An example of plume trajectories for a set of "sources" chosen by hand is shown in Fig~\ref{fig:traj_sim_radio}. A table with the positions of sources is given in the Appendix~\ref{app:plumes}. The trajectories have been integrated for $\sim 140\,{\rm Myr}$, which is approximately one rotation period around the Galactic Center near the Sun for the adopted value $\varv_{\rm rot}$.

Figure~\ref{fig:traj_sim_radio} illustrates the key morphological properties of the plumes. Naturally, the trajectories remain confined to the regions defined by the radial distance to the source from the Galactic Center, while the notable left-right asymmetry is imposed by the rotation direction of the Galaxy. All trajectories have rising parts close to the Galactic Plane, with the direction of the curvature set by the mutual position of the source and the Sun and the direction of the Galaxy rotation. This is clearly illustrated by the family of trajectories for sources located at the 5 kpc ring around GC. The morphological difference is even more pronounced for sources located at larger distances from GC (cases marked at Cygnus X and Vela in Table~\ref{tab:src}). For the Cygnus X region, the plume's apparent trajectory rises to the Galactic Poles before forming a spiral. For the Vela region, the plume is bent much earlier and follows an almost horizontal line at $|b|\sim 20-25$ degrees.  

If NPS is indeed a gaseous plume, its base should be associated with the star formation regions that are currently at Galactic longitudes $l\sim 20^\circ-40^\circ$ (see Fig.~\ref{fig:traj_sim_radio}). Incidentally, this range of longitudes corresponds to the tangential direction to the 3-5 kpc rings around the Galactic Center and indeed hosts the most active star-forming regions in the Galaxy, including a prototypical mini-starburst complex W43 (see Appendix~\ref{app:w43}). Below we discuss this scenario in more detail.

\section{Discussion}
\label{sec:dis}

Motivated by the morphological model presented above, we assume that the leading edge of NPS is located at a tangent to the 5~kpc ring. In this case, the distance of the NPS base from the Sun is $D_{\rm Sun,NPS}\approx 6.6\,{\rm kpc}$. Of course, some parts of the plumes can be closer to the Sun, while some further away. 

Given the apparent size of NPS on the sky, one can assume that its physical size (the "depth" along the line of sight) is of the same order as the distance from the Sun, i.e. $l\sim D_{\rm Sun,NPS}$. Based on that, we can estimate the hot gas density from the observed peak surface brightness $I_X$ in the 0.7-1.05~keV band. This yields a proton number density 
\begin{eqnarray}
n_p\sim 7\times 10^{-4} \left( \frac{Z}{Z_\odot} \right )^{-1/2} \left( \frac{l}{6\,{\rm kpc}} \right )^{-1/2}\,{\rm cm^{-3}}, 
\end{eqnarray}
where $Z$ is the metal abundance with respect to the Solar value \citep[the abundances relative to hydrogen from][are used here]{2009ARA&A..47..481A}. In this derivation, we assumed that the gas is in collisional ionization equilibrium with temperature $0.6\,{\rm keV}$ and used \texttt{APEC} model \citep{2012ApJ...756..128F} to predict the emissivity. The metallicity dependence in the above expression is valid for $Z/Z_\odot\gtrsim 0.1$.   The corresponding cooling time of the gas with temperature $kT\approx 0.6\,{\rm keV}$ can be estimated using the cooling function of \cite{1993ApJS...88..253S} as $t_{\rm cool}\sim  2\times 10^{9} \left( \frac{n}{10^{-3}\,{\rm cm^{-3}}} \right )^{-1} \left( \frac{Z}{Z_\odot} \right )^{-1/2} \,{\rm yr}$.

While the 3D geometry of NPS is uncertain, we assume that its volume is $\approx l^3$ (i.e. its line-of-sight extension is comparable to its transverse size) to get an estimate of the total energy (enthalpy) $E\sim l^3 (n_p+n_e) \frac{5}{2} kT\sim 2\times 10^{55} \, {\rm erg}$
for $l=6\,{\rm kpc}$.
Assuming that the integrated energy release associated with the formation of 1 solar mass of stars is $\sim 10^{49}\,{\rm erg}$, it needs $\sim 2\times 10^6$\,$M_{\odot}$ of gas to be converted into stars. Assuming the star-formation rate of $0.1\, M_\odot{\, \rm yr^{-1}}$, it takes $\sim 20$~Myr to generate enough energy for powering the entire NPS. The long cooling time and preferential accumulation of trajectories in the NPS region imply that it could be a result of cumulative contributions from many sources.

In the simplest version of the above scenario, the gas in NPS is in pressure equilibrium with the ambient medium. Assuming that the temperature of the ambient medium is $\sim 0.15\,{\rm keV}$, the regions filled with $\sim 0.6\,{\rm keV}$ plasma should have $\sim 4$ times lower density. Yet, NPS appears much brighter than the diffuse emission of the Galaxy. There are two plausible reasons for that. One is that for lines of Fe~XVII, Ne~IX, and Ne~X, the gas in the halo is simply too cool to produce a bright emission. The second reason is that the plumes are plausibly much more metal-rich than the halo gas. This is particularly important for O~VII lines, which are present in the spectrum of the halo. The gas in the plumes could easily be a factor of 10 or more metal-rich and even in the absence of shock, it can shine prominently (see Fig.~\ref{fig:spec}). In Appendix~\ref{app:abund} we argue that the plume metallicity is not much larger than the Solar one. This implicitly suggests that the abundance in the halo is $\lesssim 0.1$.

An important implication of the model is the global asymmetry of the Milky Way diffuse X-ray and radio emission. The asymmetry is present even if the distribution of plumes' sources is itself symmetric. In 3D, the plumes can form a symmetric pattern, but the direction of the Galaxy rotation produces the apparent asymmetry when viewed from the Sun's position. Some extra asymmetry could come from global gas motions in the Galactic halo \citep[e.g.][]{2023NatCo..14..781M}, e.g. caused by interaction with the satellite galaxies, but it comes on top of the "natural" East-West asymmetry. The North-South asymmetry, which is clearly visible in both radio and X-ray sky, in this model is attributed to the asymmetry of the primary sources of the plume. When the hot gas finds its way through the dense cold gas in the disk, it may preferentially go to one side rather than forming a symmetric structure on both sides.

The morphological model considered above, treated trajectories of individual gas plumes as independent, neglecting the hydrodynamic nature of the flows and, in particular, possible gas mixing instabilities that would naturally arise in such a situation. The sharpness of the NPS's outer edge as well as its overall filamentary appearance might be a manifestation of the suppression of these instabilities by the well-ordered and sufficiently strong magnetic field. 

Indeed, the synchrotron radio emission from the NPS region is known to be polarized \citep[e.g. ][ for a recent analysis]{2015ApJ...811...40S}. 
The polarization implies that the magnetic field is ordered. In the model discussed above, these structures can be especially prominent plumes located at different distances from the Galactic Center. The ordering is caused by stretching the field lines during certain phases of plume evolution and/or by the halo gas motions rather than by shocks (as illustrated in Appendix \ref{app:chimney}). 

On a more speculative side, we note that in this model the gas vented from the disk tends to accumulate high above the plane in the general direction of the Galactic Center. From this point of view, a question arises whether structures like the eROSITA bubbles \citep{2020Natur.588..227P} could be produced by the same mechanism. In this case, these structures can evolve on a longer time scale than implied by the shock-driven scenario. In fact, different regions across NPS could come from gas lumps having different ages (and different star-forming regions) and, therefore, can have different properties.  

In the proposed scenario, the NPS might be a Galactic analog of the magnetic structures directly observed to reach significant distances above the disks in many nearby galaxies, e.g. NGC~4217 \citep{2020A&A...639A.111S}. On the other hand, NPS might also be a case similar to the off-disk parts of the so-called anomalous spiral arms in NGC~4258, which are believed to be powered by the interaction of the relativistic jet with the galactic disk \citep[][ and references therein]{2023MNRAS.526..483Z}. In all these cases, we might witness signatures of the disk-halo interaction via intense feedback episodes resulting in complex multi-phase and magnetized interface region \citep[e.g. a review by][]{2015A&ARv..24....4B}.

One observational test that can falsify the plume scenario is the measurements of the non-equilibrium ionization (NEI)  signatures in the gas, which are pertinent to the shock scenario \citep[e.g., ][]{2022MNRAS.512.2034Y}. Indeed, for the gas electron density $\sim 10^{-3}\,{\rm cm^{-3}}$ and the downstream plasma velocity $\sim 200 \,{\rm km\,s^{-1}}$, the ionization parameter $\tau=n_e\times t\sim 1.5\times 10^{11} \,{\rm cm^{-3}\,s}$ is expected at a distance of $1\,{\rm kpc}$ downstream of the shock. Such scales are easily resolved for any distance to NPS (for the "nearby" shock scenario the ionization parameter is even lower at the same angular distance from the shock) and can be derived from the spectra. The complication here is that NPS does not have a simple "layered" structure that allows for clean separation of shells at different 3D distances from the edge. As a result, the lines of O~VII, O~VIII, Fe~XVII, Ne~IX, and Ne~X are always present in a proportion that is difficult to predict unless the geometry of the gas distribution is specified. We discuss the NEI scenario in a forthcoming publication. There we adopt the shock-driven scenario for NPS, consider the implications for the X-ray spectra, and compare the two scenarios.

If the model outlined above is correct, NPS and similar structures offer a possibility to probe the rotation pattern of the hot gas in the Milky Way. Future large grasp microcalorimetric missions like Line Emission Mapper \citep[LEM,][]{2022arXiv221109827K} capable of mapping the entire NPS region in the course of its all-sky survey \citep{2023arXiv231016038K} will be instrumental in measuring abundances and velocities using emission lines of elements from carbon to iron across the 0.2-2 keV energy band to narrow the range of plausible models.

\section{Conclusions}
While the North Polar Spur is usually attributed to a shock front associated with the activity of our Galactic Center or a local SNR, we discuss an alternative scenario. It posits that metal-enriched plumes rise above the disk from active star-forming regions. Interactions with the hot halo gas give these plumes the appearance of bent spirals. Their shapes sensitively depend on the rotation pattern of the hot gas above the disk. These plumes are mostly hotter than the ambient gas and can be metal-rich compared to the hot gas in the Milky Way halo. 
In this model, the NPS itself is associated with the star formation within a 3-5 kpc distance from the Galactic Center. Fainter plumes might be associated with other star-forming regions in the Galaxy.

\section*{Acknowledgments}

This work is partly based on observations with the eROSITA telescope onboard \textit{SRG} space observatory. The \textit{SRG} observatory was built by Roskosmos in the interests of the Russian Academy of Sciences represented by its Space Research Institute (IKI) in the framework of the Russian Federal Space Program, with the participation of the Deutsches Zentrum für Luft- und Raumfahrt (DLR). The eROSITA X-ray telescope was built by a consortium of German Institutes led by MPE, and supported by DLR. The \textit{SRG} spacecraft was designed, built, launched, and is operated by the Lavochkin Association and its subcontractors. The science data are downlinked via the Deep Space Network Antennae in Bear Lakes, Ussurijsk, and Baikonur, funded by Roskosmos. 

The development and construction of the eROSITA X-ray instrument was led by MPE, with contributions from the Dr. Karl Remeis Observatory Bamberg $\&$ ECAP (FAU Erlangen-Nuernberg), the University of Hamburg Observatory, the Leibniz Institute for Astrophysics Potsdam (AIP), and the Institute for Astronomy and Astrophysics of the University of Tübingen, with the support of DLR and the Max Planck Society. The Argelander Institute for Astronomy of the University of Bonn and the Ludwig Maximilians Universität Munich also participated in the science preparation for eROSITA. The eROSITA data were processed using the eSASS/NRTA software system developed by the German eROSITA consortium and analyzed using proprietary data reduction software developed by the Russian eROSITA Consortium.

IK acknowledges support by the COMPLEX project from the European Research Council (ERC) under the European Union’s Horizon 2020 research and innovation program grant agreement ERC-2019-AdG 882679.

\bibliographystyle{aa}
\bibliography{ref} 







\begin{appendix}

\section{X-ray data analysis}
\label{app:xraydata}

 Data from all four consecutive scans are combined together after filtering for the periods of enhanced solar activity, which causes a strongly elevated level of the instrumental background. For the imaging analysis, the data taken with all seven Telescope Modules (TMs) are combined, while for the spectral analysis, only five TMs protected by the on-chip filter (i.e. TMs 1-4, and 6) are used. Data reduction, filtering, vignetting-correction, and background subtraction are performed in the same way as was done in the previous studies exploring Galactic diffuse X-ray sources \citep{2021MNRAS.507..971C,2022MNRAS.509.6068K,
2023MNRAS.521.5536K,2024arXiv240117261K}.
In particular, the energy-dependent contribution of the instrumental background is modeled and subtracted based on the calibration data accumulated via observations with the "Closed Filter Wheel" configuration, while corrections for exposure time and vignetting are conducted so that the data are characterized by Field-of-View averaged response matrices.

\section{Plumes in 3D}
\label{app:plumes}
In this section, we outline two simple scenarios for plume trajectories. In the first scenario, a plume is produced by a continuous source. The plume initially moves together with stars/gas in the disk and eventually joins the motion of the halo gas. In the second scenario, a trajectory of a massless buoyant bubble is considered.

\subsection{Plumes from a moving source}
\begin{figure}
\centering
\includegraphics[angle=0,trim=2cm 0cm 4cm 0cm,clip,width=0.99\columnwidth]{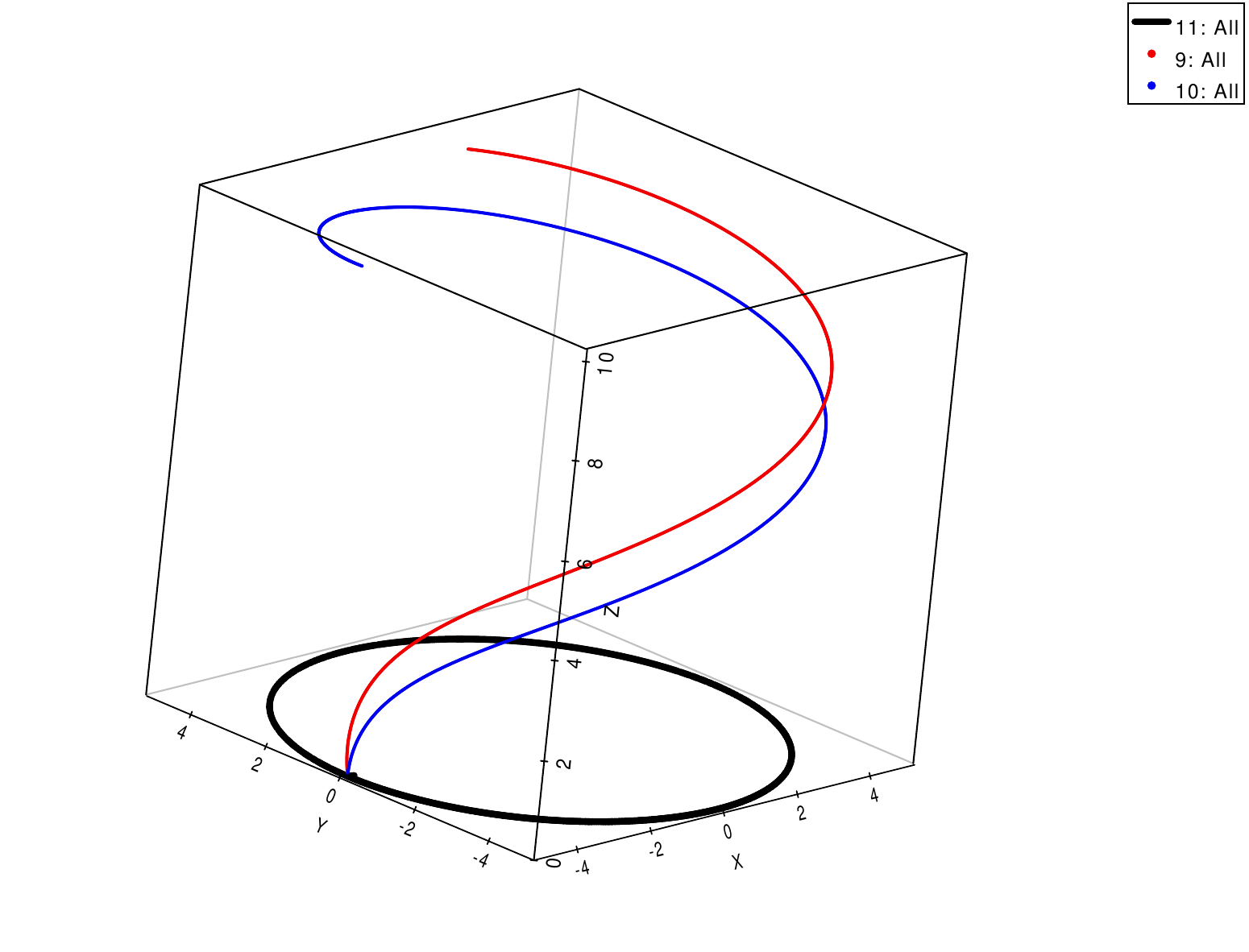}
\caption{An example of a 3D trajectory of a plume rising above the disk according to equations \ref{eq:vz} and \ref{eq:vphi} and integrated over 140~Myr. Only one side of the plume is shown. The black circle depicts a circle in the disk plane with a radius of $\sim5$~kpc. The blue line corresponds to case (a), namely the source of the plume moves together with the gas/stars in the disk plane. The red curve is the case (b) when the source of the plume, i.e. an area of active star formation, moves relative to the gas. At this distance from the GC, the difference between these two cases is not large.}
\label{fig:3d}
\end{figure}

The first case corresponds to the model described by equations \ref{eq:vz} and \ref{eq:vphi}. Fig.~\ref{fig:3d} illustrates the shapes of the "plumes" in a stationary 3D frame. The blue curve shows the case when the source of the plume moves with the same velocity as the gas in the disk. As the plume rises, it gradually joints the slowly rotating gas in the disk. The red curve shows modest modifications that appear when the source of the plume moves slower than the gas and stars in the disk. By design, the rising plumes retain the same distance from the Galactic Center in projection to the disk.   

\begin{table}[]
    \begin{tabular}{r|r|r|r|l}
    \hline
         $l$ & $b$ & $D_{\rm Earth}$ & $D_{\rm GC}$ & Comments\\
         \hline 
23$^\circ$  & 0$^\circ$ &  6   &  3.6 & \\ 
23$^\circ$  & 0$^\circ$ & 12   &  5.4 & $\sim$5-kpc-ring, far \\ 
0$^\circ$   & 0$^\circ$ &  3   &  5.3 & $\sim$5-kpc-ring, near \\ 
0$^\circ$   & 0$^\circ$ & 13   &  4.7 & $\sim$5-kpc-ring, far \\ 
332$^\circ$ & 0$^\circ$ & 10   &  4.7 & $\sim$5-kpc-ring, far \\ 
332$^\circ$ & 0$^\circ$ & 3    &  5.8 & $\sim$ 5-kpc-ring, near \\ 
82$^\circ$  & 0$^\circ$ & 1.4  &  8.2 & Cygnus region, near \\ 
82$^\circ$  & 0$^\circ$ & 3.4  &  8.5 & Cygnus region, far \\ 
262$^\circ$ & 0$^\circ$ & 1    &  8.5 & Vela region, near \\ 
262$^\circ$ & 0$^\circ$ & 5    &  10 & Vela region, far \\ 
45$^\circ$  & 0$^\circ$ & 3    &  6.5 &\\ 
45$^\circ$  & 0$^\circ$ & 9    & 6.7 &\\ 
         \hline
         \hline
    \end{tabular}
    \vspace{2mm}
    \caption{A set of "sources" used to illustrate typical trajectories of gaseous plumes in Fig.~\ref{fig:traj_sim_radio}. In the first two columns, the apparent Galactic Coordinates of the current source position are given. The next two columns give the distance (in kpc) from the Earth and the Galactic center. These positions coupled with equations \ref{eq:vz}-\ref{eq:vphi} were used to generate trajectories shown in Fig.~\ref{fig:traj_sim_radio}. }
    \label{tab:src}
\end{table}

\subsection{Trajectory of a massless buoyant bubble}
\begin{figure}
\centering
\includegraphics[angle=0,trim=2cm 0cm 3cm 0cm,clip,width=0.99\columnwidth]{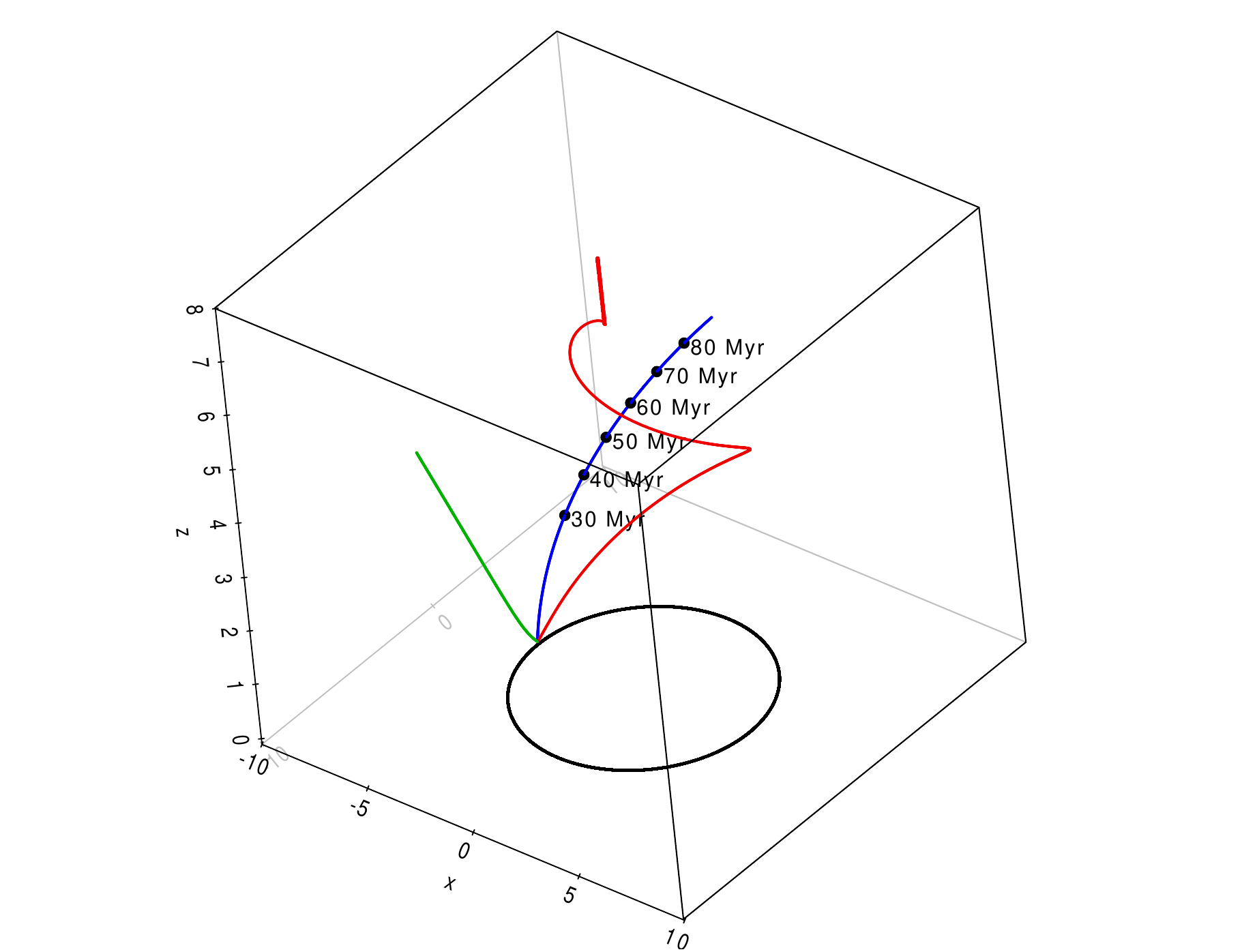}
\caption{3D trajectory of a (single) small and massless bubble rising under the action of buoyancy in a rotating but otherwise static atmosphere. The initial bubble is released at a distance of 5~kpc from the center (just above the disk plane). The green line shows the case of a non-rotating atmosphere. In this case, the bubble moves away from the Galaxy center in the disk and then switches to a more radial trajectory. The blue curve shows the case of a slowly rotating halo ($V_h\sim 100\,{\rm km\,s^{-1}}$). The bubble is now involved in rotation and motion towards larger radii. Finally, the red curve illustrates the case of fast rotation of the halo gas ($V_h\sim 200\,{\rm km\,s^{-1}}$). In this case, the centrifugal force is strong, and an inverted pressure gradient pushes the bubble closer to the rotation axis.}
\label{fig:3dbub}
\end{figure}

Here we consider the trajectory of a single (massless) bubble moving in a stratified atmosphere of the Galaxy. 
Similarly to galaxy clusters \citep{1973Natur.244...80G,2000A&A...356..788C}, the velocity of the bubble is set by the balance of pressure gradients and the drag force acting on the bubble, i.e.
\begin{eqnarray}
    u^2\frac{\vec{u}}{u} \frac{1}{2}\rho A C_d=-\nabla P V,
\end{eqnarray}
where $\vec{u}$ is the 3D velocity vector of the bubble velocity with respect to the halo gas ($u=\left | \vec{u} \right |$), $\rho$ and $P$ are the halo gas density and pressure, respectively, $A$ and $V$ are the area and volume of the bubble, and $C_d$ is the appropriate drag coefficient \citep[e.g.][]{2018MNRAS.478.4785Z}. The bubble size is assumed to be smaller than the pressure scale height and the velocity remains subsonic. For an atmosphere in hydrostatic equilibrium, $1/\rho\nabla P=-\nabla \phi$ and, therefore,   $\vec{u}$ can be directly calculated from the gravitational potential $\phi$, which in our case can include a contribution from halo gas rotation. 
To this end, we use the approximation of the MW potential from \cite{2016A&A...593A.108B} to which we add a centrifugal term ($-V_h^2\ln R$) for a cylindrical rotation, where $R$ is the distance from the MW center in the disk plane and $V_h$ is the gas rotation velocity. The rotation velocity of the gas is added to $\vec{u}$ to get the bubble velocity in the non-rotating frame. The effects of rotation on the bubble trajectory are illustrated in Fig.~\ref{fig:3dbub}. This figure shows that different configurations are possible depending on the gas rotation pattern. The three solutions shown in the figure, cover the plausible range of halo rotations. For a non-rotating halo (green curve) the bubble moves along the potential gradient created by the Milky Way mass distribution. The blue and red lines cover the cases of moderate ($100\,{\rm km\,s^{-1}}$)
and extreme ($200\,{\rm km\,s^{-1}}$) halo rotation speed. In particular, the case of the halo rotation velocity with $\sim 180\,{\rm km\,s^{-1}}$ \citep{2016ApJ...822...21H} should be more close to the red curve. Of course, the rotation pattern can be much more complicated than the cylindrical pattern considered here.

\begin{figure}
\centering
\includegraphics[angle=0,trim=2cm 0cm 3cm 0cm,clip,width=0.99\columnwidth]{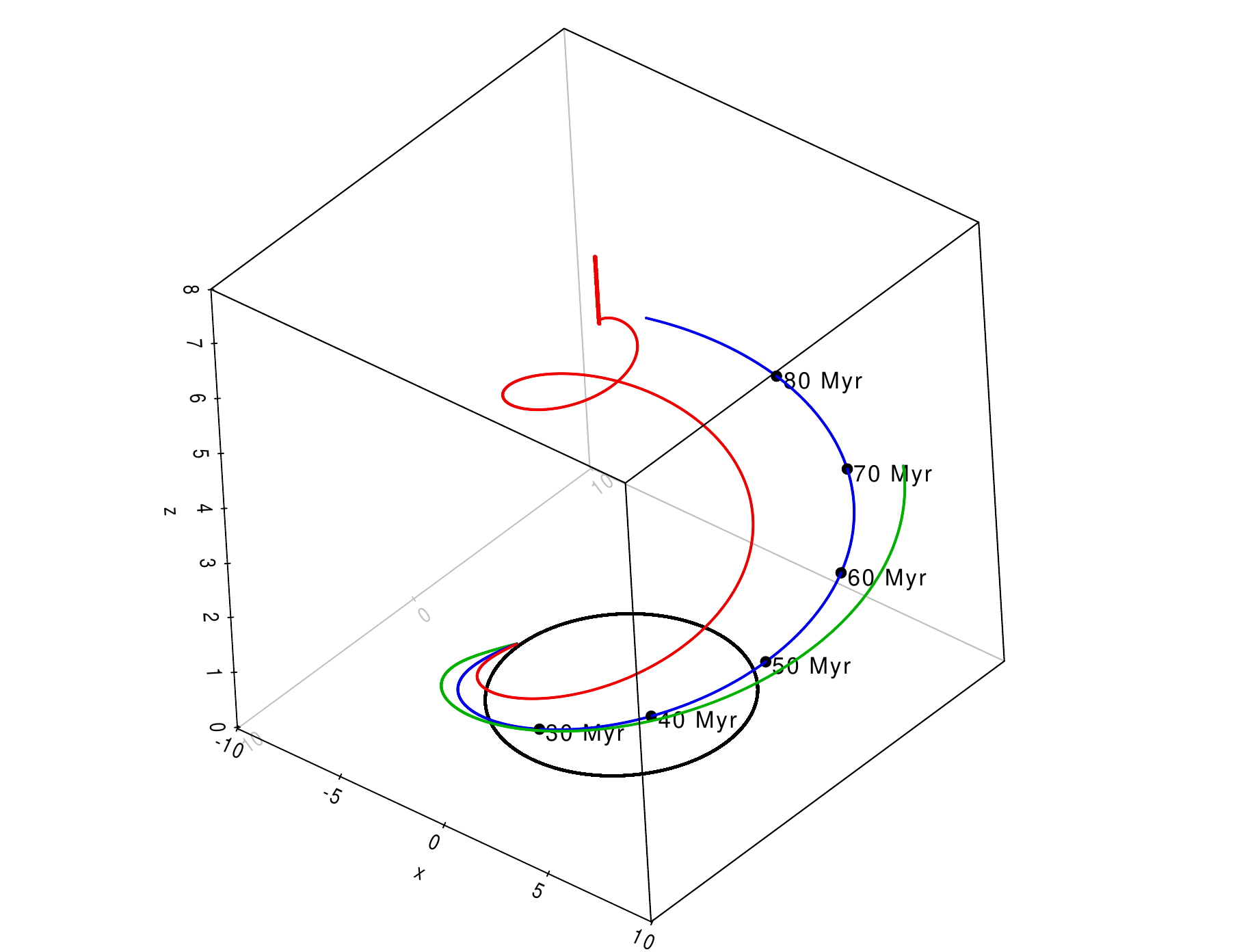}
\caption{3D trajectory of a sequence of bubbles (i.e., a plume). Individual bubbles follow the trajectories shown in Fig.~\ref{fig:3dbub}. The source of the bubbles moves with the same velocity as stars in the disk. Overall, plume trajectories are qualitatively similar to those shown in Fig.~\ref{fig:3dplu}. }
\label{fig:3dplu}
\end{figure}

We also emphasize that Fig.~\ref{fig:3dbub} (unlike Fig.~\ref{fig:3d}) shows trajectories of individual "bubbles" rather than "plumes" that are formed by many bubbles released at different times. Assuming that the source of the bubbles is moving together with stars in the disk, the plume trajectories (a combination of bubbles released at different times) are shown in Fig.~\ref{fig:3dplu}. Qualitatively, they resemble the plumes shown in Fig.~\ref{fig:3d}, except for the presence of radial migration.

\section{Enhanced metallicity of plumes}
\label{app:abund}

The enhanced metallicity of the NPS hot material enriched by supernovae is important for the proposed scenario. 
The thermal energy of the NPS $E_{\rm th} \sim 2 \times 10^{55}$\,erg (see \S\ref{sec:dis}) suggests the explosion of $N_{\rm sn}\sim 2\times 10^4$ 
  core-collapse supernovae (CCSN) with a typical kinetic energy of $10^{51}$\,erg (lifetime of a star-forming region is short for SNe~Ia to explode). 
We rely on two sets of oxygen nucleosynthesis calculations  for CCSN 
 progenitors in the range of 11 - 40$M_{\odot}$ by \cite{WW_1995}
 (WW95) and 
\cite{Kobayashi_2006} 
  (K06).
In K06, the data on 11$M_{\odot}$ progenitors are lacking. We, therefore, adopt the ejected mass of oxygen from WW95 for this particular progenitor.  The abundance of other elements relative to oxygen is assumed to follow (approximately) the solar composition. 

 The average mass of ejected oxygen per CCSN ($m_{\rm O}$) is inferred assuming Salpeter initial mass function   $dN/dm \propto m^{-2.35}$ in the mass range of 0.1 -- 100$M_{\odot}$. We find comparable values of $m_{\rm O}$, 2.2$M_{\odot}$\ and 2.6$M_{\odot}$,  
  for WW95 and K06 data, respectively.
The average $m_{\rm O}=2.4$$M_{\odot}$\ multiplied by the CCSN number $N_{sn}$  provides us with the total amount of oxygen synthesized by supernovae of the star-forming region $M_{\rm O} \approx 5\times 10^4$$M_{\odot}$.
The synthesized oxygen mixed with the gas of the solar composition (solar abundance $X(\mbox{O})_\odot\ = 0.01$) produces enhanced overall metallicity. 
The total mass of the mixture ($M_{\rm hot}$) is fixed by the thermal energy of the X-ray-emitting gas 
 $E_{\rm th} = (5/2)kT_{\rm x}M_{\rm hot}/(\mu m_p)$. 
For the average molecular weight $\mu = 0.61$ and temperature $kT_{\rm x} =0.6$\,keV one obtains  
 $M_{\rm hot} = 4\times10^6$$M_{\odot}$. 
The expected oxygen abundance of the NPS hot gas is then $X(\mbox{O}) = X(\mbox{O})_\odot\ + M_{\rm O}/M_{\rm hot} \approx 0.02$, twice the solar value.

The effect of the enhanced oxygen abundance of the NPS can be expressed formally as $X(\mbox{O}) = X(\mbox{O})_{\odot}(1 + y)$ with 
 the factor $y$ responsible for the synthesized oxygen and estimated for NPS to be about unity.
 
 Let the abundance of a certain element (El) relative to oxygen produced by CCSNe with respect to the solar abundance is $\phi$=(El/O)/(El/O)$_{\odot}$. 
The overabundance factor of this element with respect to solar in the NPS material, assuming the initial abundance in the star-forming region to be solar, is then  $f = 1 + y\phi$. 

The relative abundances  $\phi$ of key elements, viz. 
C, N, Ne, Mg, Si, S, and Fe, generated by the CCSNe, can be derived from spectroscopic stellar data of [El/Fe] = log(El/Fe) - log(El/Fe)$_{\odot}$ for metal-poor stars ([Fe/H] < -2), in which metallicity domain the nucleosynthesis by CCSNe dominates. We rely on  [El/Fe] vs. [Fe/H] data compiled by Kobayashi et al. (2020). 
 For C and N the scatter of [El/Fe] values relative to the average for different stars is very large, of $\sim$0.5 dex; in other cases, the scatter is of 0.2--0.3 dex.
Anyway, for each element we obtain an average value of [El/Fe]  with a typical error of 0.1 dex, which is converted to 
 $\phi$ with a relative error of $\approx 20$\%. In the case of Ne, lacking stellar data, we rely on the theoretical prediction that CCSNe nucleosynthesis results in the solar ratio of Ne/O \citep{2020ApJ...900..179K}. 
For the adopted value of $y=1$, the estimated overabundance factors in the NPS material  
turn out to be $f = 1.5$ for C, N, and S,  $f = 1.6$ for
 Mg and Si,  $f = 2$ for O and Ne, and $f = 1.25$ in the case of Fe -- all factors with a relative error of approximately 20\%.

\section{The chimney model}
\label{app:chimney}

The clustering of core-collapse supernovae produced at the end of the evolution of massive stars in OB associations and compact clusters has a profound effect on the interstellar medium 
and the galactic ecology \citep{1988ApJ...324..776M,1990ApJ...354..483H,2000MNRAS.315..479D,2020ApJ...900...61K}. Multiple correlated supernovae and the powerful winds of OB stars create superbubbles and eventually (for powerful enough systems) produce the chimney-type structures \citep{1989ApJ...345..372N}. We consider such superbubbles as potential sources of the plumes. 
 To study the effect of the galactic magnetic field on the superbubble breakthrough from the disk to halo \citet{1998MNRAS.298..797T} performed 3D MHD simulations for different assumptions on the large-scale magnetic field structure. The magnetic field with the strength of $\sim$ 5 $\mu$G with a broad $\sim$ kpc scale height can confine a superbubble of a modest kinetic luminosity 3$\times 10^{37} \ergs$  for about 20 Myr within the scale height of $\mid z \mid \sim$ 300 pc from the disk, while in a model with the field that follows the scaling $B \propto \rho^{1/2}$ (with the mid-plane field magnitude of 5 $\mu$G) the superbubble blows out to the halo. The evolution of superbubbles in density stratified disks that blow out into galactic halo may be a subject of Rayleigh-Taylor instabilities \citep{2013A&A...557A.140B,2022MNRAS.509..716S}. 
 A parsec-resolution multi-phase simulation of the local star-forming galactic disk with the account for SNe feedback effects revealed that the hot galactic outflows (with gas temperatures above 10$^6$ K) may carry about 10\%-20\% of the energy and 30\%-60\% of the metal mass injected by supernovae \citep{2020ApJ...900...61K}. These values are broadly consistent with the energy requirements needed to produce the X-ray plumes discussed in the text.  
 
 Non-thermal particles accelerated by shocks from SNe and stellar winds at the active phase of the evolution of superbubble which is about 10 Myrs \citep{2014A&ARv..22...77B}. Then relativistic particles will be blown out to the low halo with the frozen-in magnetic fields of the chimney-type plasma outflow. Relativistic electrons of energy $\sim$ 100 GeV would have in the halo a lifetime of about 10 Myrs due to synchrotron - Compton losses and will radiate in magnetic fields of a few $\mu$G at frequencies of $\lsim$ 50 GHz. The synchrotron-Compton cooling time for the relativistic electron of energy ${\cal{E}_{\rm GeV}}$ (measured in GeV) in the galactic halo can be estimated as 
 \begin{eqnarray}
t_{\rm syn}\sim  10^{9} \frac{{\cal{E}_{\rm GeV}}^{-1}}{ 1 + 0.1 B_{\rm \mu G}^2} \,{\rm yr}. 
\end{eqnarray}
\begin{figure}
\centering
\includegraphics[angle=0,trim=2cm 2cm 0cm 0cm,clip,width=0.99\columnwidth]{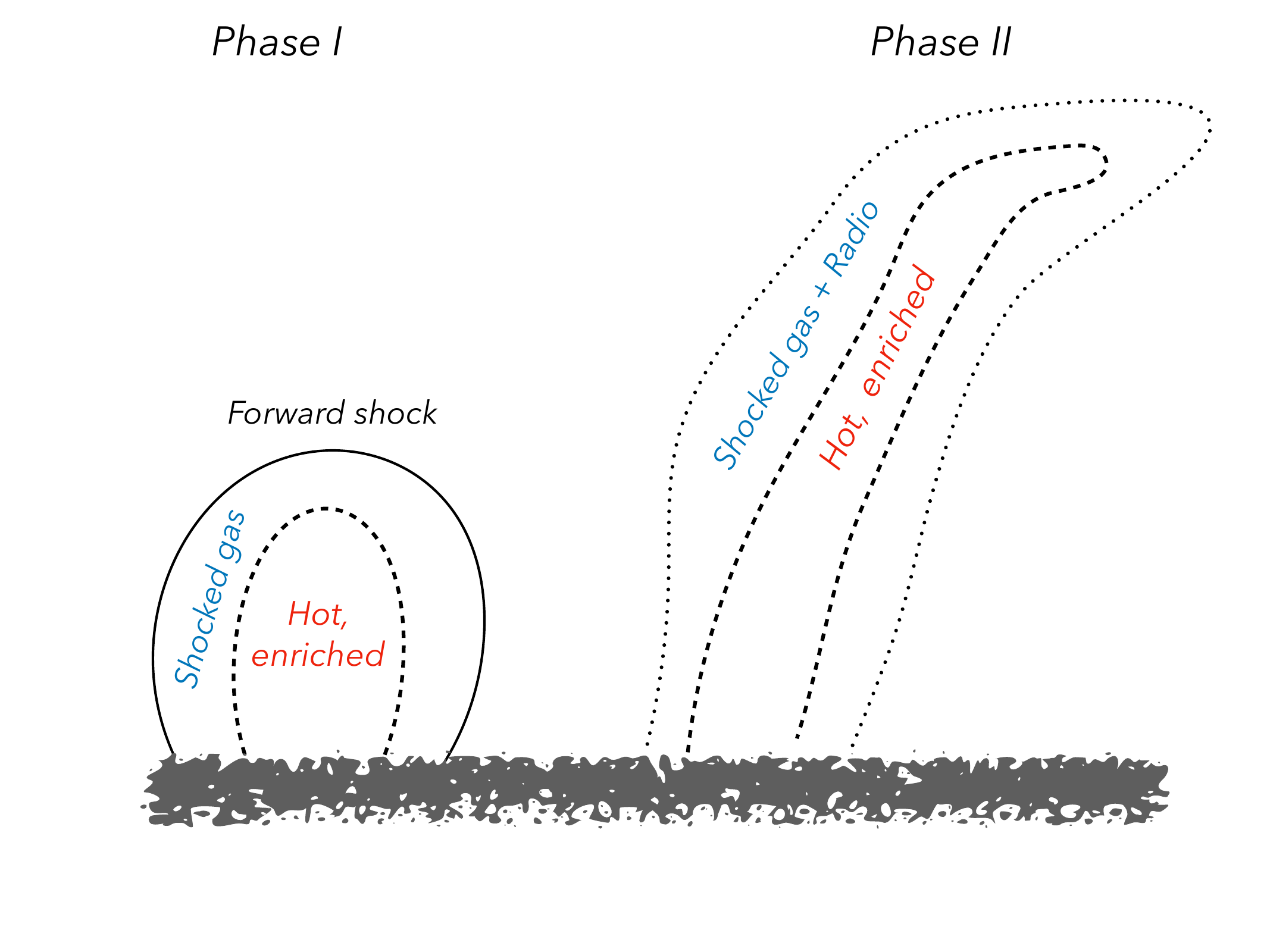}
\caption{Possible formation scenario of a NPS-like structure. During the initial phase after the breakout (left), a shock driven by the hot and metal-rich gas of a superbubble propagates through the ambient gas. At later stages (right) the gas is stretched and bent into a plume but possibly maintains the layered structure. }
\label{fig:sketch}
\end{figure}

\section{The role of W43}
\label{app:w43}
Below, we discuss in brief the mini-starburst complex W~43 as a prototype 
powerhouse located in the Galactic molecular ring assuming that an ensemble of such complexes 
may supply the plumes over a long period of $\sim$ hundred million years. 
The molecular complex W~43 with the estimated 
mass $\sim 7 \times 10^6\, \Msun$ is presumably located at the connecting point
of the Scutum-Centaurus Galactic arm and the Galactic bar \citep{2011A&A...529A..41N}. W 43 excited a giant H II region dubbed NGC 3603. The equivalent diameter of the coherent complex of molecular clouds is $\sim 140$ pc and it is surrounded by an atomic gas envelope of about two times larger diameter. \citet{2011A&A...529A..41N} noted that the transition from circular to elliptical orbits
in the spiral arm and bar potentials might cause high-velocity streams and thus efficient star formation episodes of current rate at least $\sim 0.01\, \Msun$~yr$^{-1}$ as estimated from 8 $\mu$m luminosity measurements but it is likely increasing to $\sim 0.1\, \Msun$~yr$^{-1}$. On longer time scales $\sim$ 100 Myrs of interest for this paper multiple episodes of high star formation rate can be expected from colliding molecular clouds entrained by the large scale streams.   

The giant HII region NGC~3603 which is associated with the complex produces the Lyman continuum photon rate of about  $4 \times 10^{51}$ photons s$^{-1}$ is the most powerful HII region in the Galaxy \citep{2024ApJ...963...55D}. A compact dense cluster of young massive stars HD 97950 (also known as W~43 cluster)  powering NGC~3603 has an estimated dynamic stellar mass of 18,000~$\Msun$ including   
more than 70 O and 3 WR stars. The distance to the powerful starburst region NGC~3603 
of 7.2 $\pm$ 0.1 kpc was estimated by \citet{2019MNRAS.486.1034D} from GAIA DR2 stellar parallax measurements (see also \citet{2024ApJ...963...55D} for a recent discussion). From the VLBA trigonometric parallax measurements of masers  toward W43 \citet{2014ApJ...781...89Z} derived a distance of 5.49$_{-0.34}^{+0.39}$ kpc to the complex \citep[see also][for another distance estimate of $\sim 4.8\,{\rm kpc}$]{2022ApJS..262...42L}.  The origin and ages of 
molecular clumps and stellar population in NGC 3603 are still the subject of debate \citet{2024ApJ...963...55D}. ALMA observations revealed that the core mass function in W43-MM1 is much shallower in the high-mass range than the standard initial mass function \citep{2018NatAs...2..478M}.  
\citet{2014ApJ...780...36F,2021PASJ...73S.405F} discussed the formation of cluster W 43  due to a collision of two molecular clouds about 1 Myr ago. This age is consistent with that of the highest mass stellar population in the cluster. 
The suggestion by \citet{2011A&A...529A..41N} about the origin of the star-forming complex W 43/NGC 3603 due to the collision of molecular clouds following their circular orbits in the Scutum-Centaurus arm and the clouds at the elliptical orbits of the Galactic bar makes the complex W 43 as well as other complexes along the Molecular Ring where the Galactic arms meet the bar to be favorable locations for the intense star formation episodes in the past.   
 
Very high energy emission source HESS J1848-018 \citep{2018A&A...612A...1H} can possibly be associated with W 43. From  Fermi-LAT detection of extended high energy gamma-ray source source in the direction of W 43 \citet{2020A&A...640A..60Y} estimated the total cosmic ray energy within the source to explain the detected flux to be $(2.3 \pm 0.3) \times 10^{48}$ erg.


\end{appendix}


\label{lastpage}
\end{document}